\documentclass[prb,twocolumn,showpacs]{revtex4-1}
\usepackage{amsfonts,amsmath,mathrsfs,epsfig,amsbsy,bm,verbatim}

\begin{document}

\title{Density of states of disordered topological superconductor-semiconductor hybrid nanowires}
\author{Jay D. Sau$^1$}
\author{S. Das Sarma$^2$}

\affiliation{$^1$Department of Physics, Harvard University, Cambridge, MA 02138
\\
$^2$Condensed Matter Theory Center and Joint Quantum Institute, Department
of Physics, University of Maryland, College Park, Maryland 20742-4111, USA.
}

\begin{abstract}
Using Bogoliubov-de Gennes (BdG) equations we numerically
 calculate the disorder averaged density of states 
of disordered semiconductor nanowires driven into a putative
 topological $p$-wave superconducting phase 
by spin-orbit coupling,  Zeeman spin splitting and $s$-wave
superconducting proximity effect induced by a nearby 
superconductor. Comparing with the corresponding theoretical 
 self-consistent Born approximation (SCBA) results treating 
disorder effects, we comment on the topological phase diagram of 
the system in the presence of increasing disorder. Although 
disorder strongly suppresses the zero-bias peak (ZBP) associated 
with the Majorana zero mode, we find some clear remnant of a ZBP 
even when the topological gap has essentially vanished in the 
SCBA theory because of disorder. We explicitly compare effects of disorder 
on the numerical density of states in the topological and trivial phases.
\end{abstract}

\maketitle

\section{Introduction}
The theoretical prediction \cite{1,2,3,4,5,6} that the combination 
of spin-orbit coupling, Zeeman spin splitting, and ordinary 
$s$-wave superconductivity could lead to an effective topological 
superconducting phase under appropriate (and experimentally 
achievable) conditions has led to an explosion of theoretical 
and experimental activities \cite{6} in semiconductor nanowires 
(InSb or InAs) in proximity to a superconductor (NbTi or Al) 
in the presence of an external magnetic field.
The experimental finding \cite{7} of a ZBP, in precise agreement with the 
theoretical predictions in the differential tunneling conductance of an 
InSb nanowire (in contact with a NbTiN superconducting substrate) at 
a finite external magnetic field ($B\sim 0.1-1$ T), followed by independent 
corroborative observation \cite{8,9,10,11} of such ZBP both in InSb and InAs 
nanowires in contact with superconducting Nb and Al by several groups, 
has created excitement in the condensed matter physics community as well as 
the broader scientific community as perhaps the first direct evidence supporting 
the existence of the exotic, the elusive, and the emergent unparied Majorana bound state 
in solids. Such excitement has invariably been followed by a wave of skepticism as one 
would expect in a healthy and active scientific discipline with questions 
ranging all the way from whether such ZBP could arise from other (i.e. non-Majorana)
origin to whether all aspects of the observed experimental phenomenology 
are consistent with the putative theoretical predictions on the topological 
superconductivity underlying the existence of the Majorana mode. 

One particular issue, which is also the subject of the current work, 
attracting a great deal of theoretical attention \cite{12,13,14,15,16,17,18,19,20,21,22,25,26} 
is the role of disorder in the Majorana physics of superconductor-semiconductor hybrid 
structures. Disorder plays a key role in the Majorana physics because the underlying 
topological superconducting phase hosting the Majorana mode (at defect sites) 
is essentially an effective spinless p-wave superconductor \cite{1,2,3,4,5}, 
which, unlike its $s$-wave counterpart, is not immune to non-magnetic elastic disorder 
(i.e. spin-independent momentum scattering) as was already known ten years ago \cite{23}. 
Thus, even the simplest kind of disorder, namely zero-range random non-magnetic point 
elastic scatterers in the wire, could strongly affect the topological superconductivity 
in contrast to ordinary $s$-wave superconductivity which is immune to non-magnetic 
disorder, and the associated Majorana bound states by suppressing the (topological) 
superconducting gap \cite{13} and/or creating Andreev bound states in the superconducting 
gap near zero-energy \cite{14} complicating the observation and the interpretation 
of the ZBP. We do, however, mention that elastic disorder or momentum scattering in the superconductor 
itself, no matter how strong (as long as it does not destroy the superconductor),
 does not affect the topological superconducting phase in the semicondutor. \cite{21}
In addition, it has recently been emphasized that elastic disorder by itself 
could create a zero-bias peak (essentially, an anti-localization peak associated with the  
disorder-induced quantum interference) in the non-topological phase in the presence 
of spin-orbit coupling and Zeeman splitting. Since the precise topological quantum 
critical point (as a function of the applied magnetic field) separating the topological 
and the trivial superconducting phase is in general not known in the experiments, \cite{7,8,9,10,11,12} one cannot 
be absolutely sure that the observed ZBP is indeed a Majorana bound state (MBS) 
signature in the topological phase and not a trivial anti-localization peak in the non-topological 
phase.

In the current work we consider disorder effects on MBS physics by directly calculating the density of states (DOS) 
of finite disordered nanowires in the presence of proximity-induced superconductivity taking into account 
spin-orbit coupling and spin splitting arising respectively from Rashba and Zeeman effects in the 
wire. We use random uncorrelated point scatterers with $\delta$-function potential in the wire to represent
the elastic disorder. The theory follows the standard prescription \cite{24} of an exact diagonalization 
of the BdG equations in a minimal tight-binding model including superconducting pairing, Rashba spin-orbit 
coupling, and Zeeman splitting in the Hamiltonian. The diagonalization of the 
discretized tight-binding Hamiltonian leads to the exact eigenstates of the system, which then 
immediately give the density of states (theoretical details are available in Ref.~\onlinecite{24} and are not 
repeated here.) Comparison with theory in the presence of disorder neccessitates the ensemble averaging over many different impurity 
configurations since each disorder configuration produces its own unique result with random impurity-induced 
delta-function peaks in the superconducting gap.

\section{DOS in the disordered topological phase}
In this section, we consider the effect of disorder on the SM/SC system starting from the topological phase in the clean 
limit.
In Fig. 1 (with 8 panels, each representing a different strength of disorder keeping all other parameters fixed), 
we show our numerical DOS as a function of energy (E) for different relevant parameter sets in the topological phase of the system. We 
show both the ensemble averaged DOS using many-impurity configurations(but using the same disorder strength, i.e. 
the same impurity density and potential strength, changing only the random localitions of the impurities along the wire) 
and the typical DOS for a single impurity configuration in each case. Each panel corresponds to a specific disorder strength 
(i.e. a fixed impurity density) and shows results for three different lengths of the 
nanowire. 

Since our calculations follow either Ref.~\onlinecite{24} for the exact numerical treatment of Ref.~\onlinecite{13} for the SCBA theory, we refer 
the reader to those references for the technical details, which are actually pretty standard. \cite{19,20,21,22}

\begin{figure}[tbp]
\begin{center}
\includegraphics[width=0.7\textwidth,angle=270]{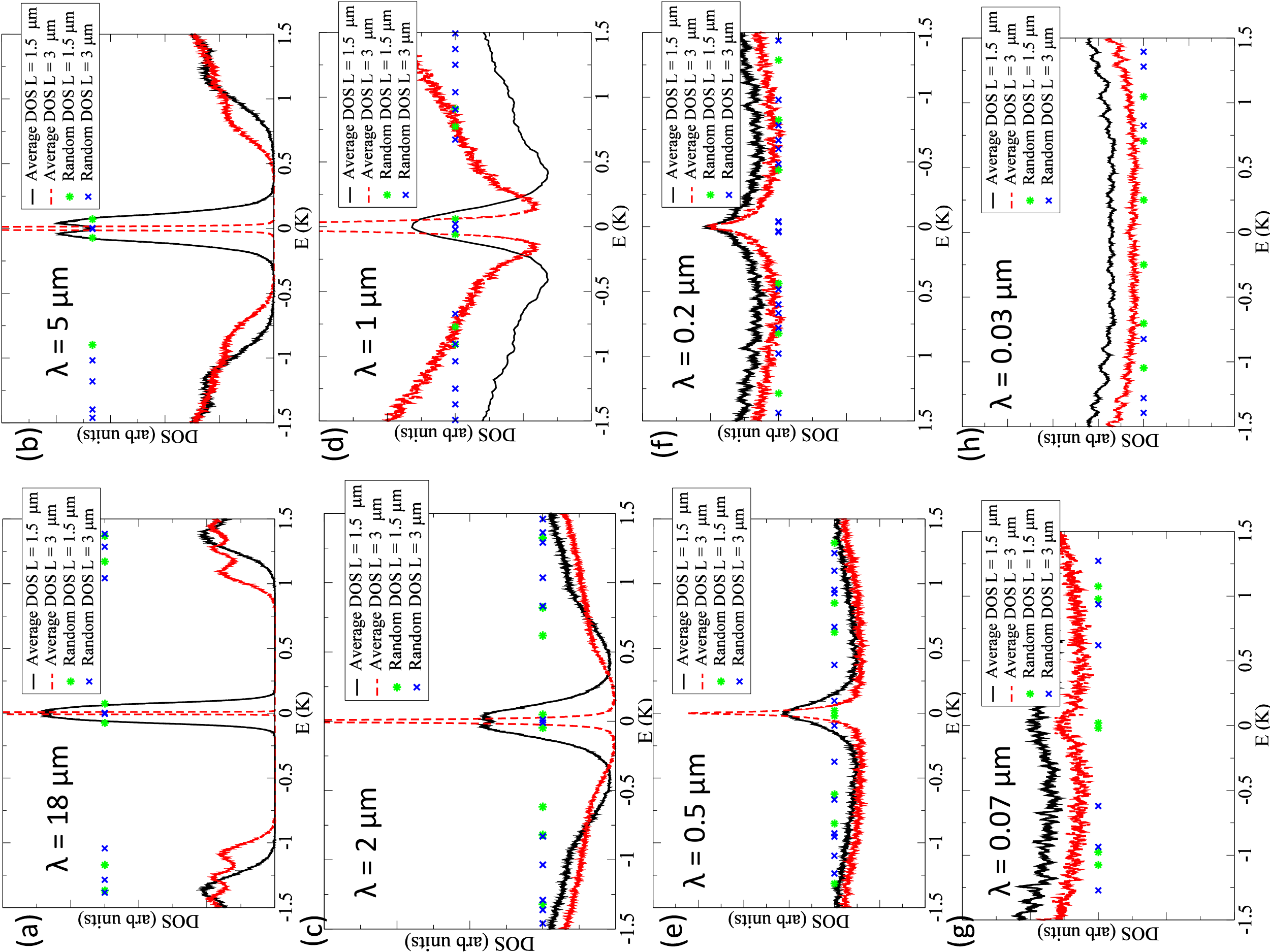}
\end{center}
\caption{Disorder averaged density of states in the topological phase for the semiconductor nanowire in a magnetic field with Zeeman splitting 
 $V_Z=5\,$K, proximity-induced pairing potential of amplitude  $\Delta_0=3\,$K, Rashba spin-orbit coupling strength $\alpha_R=0.3\, eV-\AA$. For 
this choice of parameters the clean quasiparticle-gap $\Delta=1.3\,$K and coherence length $\xi=0.3\,\mu m$. The difference panels (a-h)
correspond to different disorder strengths characterized by $E_s=\hbar/2\tau= 10, 50, 100, 200, 460, 1300, 3300, 7300$ mK.
 The corresponding mean-free paths are $\lambda$ are in the panels and the self-consistent Born gaps $E_{SCB}= 1.2, 1.1, 0.9, 0.8, 0.13, 0, 0, 0$K.}
\label{Fig1}
\end{figure}

The superconductor (SC)-semiconductor (SM) hybrid nanowire structure is characterized by a large number of independent parameters, both for 
the actual experimental laboratory systems and for our minimal theoretical model. The minimal set of parameters necessary to describe the system
 are the  proximity-induced SC gap ($\Delta_0$) in the SM, the parent SC pairing potential ($\Delta_s$), the Rashba spin-orbit coupling ($\alpha_R$),
the spin splitting $(V_Z)$, the chemical potential ($\mu$), the SM effective mass $(m)$ which defines the SM tight-binding hopping 
parameter, the nanowire length $(L)$, and disorder (which we take to be the uncorrelated random white noise potential  associated with randomly 
located $\delta$-function in real space spin-independent scattering centers). There are a few additional physically important parameters 
which are, however, not independent parameters of the theory: the coherence length in the nanowire $(\xi)$, the SM mean-free path $(\lambda)$
due to disorder, and the number of occupied subbands (i.e. transverse quantized levels) in the nanowire which we take to be one 
throughout assuming the system to be in the one-dimensional limit. (The single sub-band approximation, made entirely for the convenience 
of keeping the number of parameters in the model to be tractable, is a nonessential approximation, and our qualitative results should be 
completely independent of this approximation.) In addition, there is an independent parameter defining the hopping amplitude across the 
SC/SM interface which controls the proximity-induced SC pairing gap $\Delta_0$ in the SM in terms of the parent gap $\Delta_s$ 
in the SC. Finally, the proximity gap ($\Delta$) in the SM in the presence of spin-orbit coupling and finite Zeeman splitting is reduced 
from $\Delta_0$ in a known manner. All the details for modeling the disorder 
are given in Ref.~\onlinecite{13} where SCBA was used in contrast to 
our exact numerical diagonalization in the current work following 
Ref.~\onlinecite{24}. One specific goal of our current work is to compare the analytic and simple SCBA theory \cite{13} with the exact tight-binding numerical 
analysis to test the limits of validity and the applicability of the SCBA theory which, being analytic, can be used rather easily.

We choose parameters approximately consistent with the InSb/Nb systems studied 
experimentally in \cite{7}. These are : $\Delta_0=3$K, $\alpha_R=0.3 eV-\AA$ (corresponding to an effective spin-orbit coupling strength $m^*\alpha_R^2=2.5 K$). Since out interest is in disorder effects, we focus on a range of magnetic fields with a spin splitting $V_Z\sim 5$K (we vary it in a few cases 
only to change the proximity to the topological quantum phase transition point, separating the topological and the trivial SC phase). Given that the condition for 
the topological SC phase to be realized in the SM is given by 
\cite{2} $V_Z>\Delta^2+\mu^2$, where $\Delta$ is the actual 
induced gap in the SM, the system should be in the MBS carrying 
topological phase for $\Delta<5$K (since $\mu=0$). Given that 
$\Delta<\Delta_0=3$K, with the reduction of $\Delta$ below $\Delta_0$ arising from the existence of $V_Z\neq 0$, our system is deep in the topological 
phase for the results shown in Fig.1 since $V_Z(=5K)\gg \Delta(=1.3K)$ and $\mu=0$.

Each panel in Fig. 1 corresponds to a different disorder strength 
in the system, characterized by the corresponding level broadening $E_s=\hbar/2\tau$ 
(where $\tau$ is the scattering time -- $\tau=\infty$ in the absence of 
disorder) or equivalently the mean free path $\lambda=v_F\tau$ (where $v_F$ 
is the fermi velocity), both calculated in the SCBA according to Ref. 
\onlinecite{13} for the given disorder in the wire. In each panel 
(and for each disorder) we show our DOS numerical results for two distinct 
wire lengths $L=1.5\,\mu m$ and $L=3\,\mu m$. In each case, we show both the 
ensemble averaged DOS results using an averaging over many $(> 1000)$ random 
impurity configurations (keeping $E_s,\lambda$ etc fixed) and the result 
for a typical single impurity configuration (the distinct crosses or dots 
denoting delta functions for the DOS at the value of energy ). 
We emphasize that for an infinitely long wire ($L\gg \xi\approx 0.5\,\mu m$ 
for our case) the DOS in the absence of disorder will vanish  
throughout the gap ($\pm 1.3$ K in our case) with a $\delta$-function peak  
at $E=0$ associated with the MBS at the wire edges.

To characterize the disorder strength for the results in Fig. 1 (with panels 
(a) to (h) with increasing disorder keeping all other parameters fixed ), 
we use SCBA for this problem which was developed by us in Ref. \onlinecite{13}.
The SCBA theory provides us with the SC gap in the topological phase for a 
given disorder strength, allowing us to compare our direct (and exact) numerical 
calculation in the presence of disorder with the SCBA theory. We show the 
calculated SCBA gap in each case in the figure captions for the sake of direct 
comparison with the exact results in the figures. (The SCBA theory is obviously 
an ensemble averaged theory for the infinite system and does not depend on $L$.)
It is clear from the result of Fig. 1 that the analytical SCBA theory of Ref.
~\onlinecite{13} are in excellent qualitative agreement 
with the exact ensemble-averaged numerical results for the DOS even for 
$E_s\approx \Delta$, where the topological gap essentially vanishes 
(panel (e) in Fig. 1) both according to the SCBA (i.e. $E_{SCB}=0$) and 
in our numerical results.
While the disorder averaged DOS calculated by exact diagonalization 
does not stricly vanish inside the gap, 
the gap calculated with the SCBA can be identified with the peaks in the DOS 
at the edges of the gap. The closing of the gap within the SCBA coincides 
with the disappearance of the dips in the DOS around zero-energy.
 The DOS peak at $E=0$ associated with the MBS is 
continuously suppressed with increasing disorder, but quite amazingly 
there is a discernible DOS peak at $E=0$ even for $E_s(=3.3K)\gg \Delta(=1.3K)$
 where $E_{SCB}=0$, and at best the topological superconductivity is 
gapless.

The remarkable result, which is quite apparent in our Figs. 1 (e)-(g), 
is that the MBS peak of the DOS at $E=0$ is actually very robust to 
disorder and survives disorder strength substantially larger than that (typically $E_s\sim \Delta$) 
destroying the induced 
superconducting gap $\Delta$. Thus, in Figs. 1(e)-(g), although the 
SCBA theory and our exact numerical results both show the system to be 
gapless with $E_s>\Delta$, the DOS peak at $E=0$ associated with the MBS
persists until $E_s\gg \Delta$ as in Fig. 1(h) where $E_s=7 K(\gg \Delta=1.3 \, K)$.
 It is not only that the MBS feature in the ensemble averaged DOS 
survives up to very strong disorder (e.g. the mean free path $\lambda=0.5\,\mu m,0.2\,\mu m$ and 
$0.06\,\mu m$ respectively in Figs. 1(e)-(g) which are smaller than the 
wire lengths of $L=1.5\,\mu m$ and $3\,\mu m$ used in our numerical work),
the typical DOS for single random impurity configuration also shows peaks at 
$E=0$ as can clearly be seen in Figs. 1(e)-(g) [ and as well as in Figs. 
1(a)-(d)], but not in Fig. 1(h) where the very large disorder strength 
($\lambda=0.02\,\mu m$) suppresses both the ensemble averaged MBS peak as 
well as the single configuration peak at $E=0$. The survival of the zero-energy DOS 
peak well above the point where the SC gap is completely suppressed by disorder 
is an important new result of our exact numerical work directly establishing the 
possible theoretical existence of a gapless topological SC phase.

\subsection{Griffiths effects}
The origin of the $E=0$ peak in the DOS in the strongly disordered case 
can be related to the Griffiths effect previously considered for the spinless 
$p$-wave superconductor \cite{23}. The Griffiths effect in the semiconductor nanowire
at finite $V_Z$ arises from the disorder-induced
 variation of the chemical potential, which can lead to a transition from a topological phase 
to a non-topological phase and vice-versa. The variation of the effective chemical potential 
can lead to domain-walls between topological and 
non-topological regions each of which would support a local zero-energy MBS. 
Since each region is of finite extent, the MBSs are a finite distance apart and 
split into  conventional states with a non-zero energy. In fact, by carefully 
considering the distribution of the distances between the MBSs, it has been shown \cite{23} 
that this splitting typically leads to a singular peak in the DOS at $E=0$, 
which diverges as $E\rightarrow 0$ as a power-law in $E$. The topological phase is of course 
characterized by true zero-energy edge states, whose energy is exponentially small in the 
length of the system $L$. Thus, in the long wire limit, the topological phase and the non-topological 
phase both have power-law divergent DOS due to the Griffiths effects, but the topological phase 
has a pair of zero-modes exponentially close to zero energy \cite{12}. This distinction 
between exponential versus power-law in the length seems to appear in the DOS plotted in Fig 1.
At small disorder, there is a sharp peak which is very weakly split. At larger disorder (panels (e,f,g,h))
, which is approximately when the gap vanishes i.e. $E_{SCBA}\rightarrow 0$, the $E\sim 0$ peak in the 
DOS is broadened.

For finite length systems, the sharp transition between the 
topological and non-topological phase at finite disorder becomes a crossover.
While the topological invariant can be computed even for a disordered, but strictly infinite system \cite{12,25,26}, 
we have not done so in the present work because we restrict ourselves to systems 
of wire lengths comparable to the experimental systems. For such finite wires the distinction between the topological 
and non-topological phase is not sharp. In fact, the zero-modes arising from the Griffiths effect 
are themselves MBSs. The MBSs characterizing the topological phase are special only in the 
sense that they occur near the ends of a finite system and are therefore separated from other low-energy 
MBSs by a distance of the order of the length of the wire. This can also occur from the Griffiths 
effect in some realizations of disorder, even in the non-topological parameter regime. 
Therefore, the Griffiths singularities seen in the disordered Fig. 1 indicate the 
 presence of several low-lying MBSs in the spectrum of the finite wire. 
For very large disorder the system becomes completely non-topological (and gapless)
 as in Fig. 1(h).

There are three important messages following from our DOS results for the 
topological phase in the presence of disorder: (1) the zero-energy DOS
peak associated with the MBS is strongly suppressed by disorder; (2) SCBA
is an excellent quantitative approximation for calculating the SC gap in the 
topological phase including effects of disorder; and (3) most importantly, 
the zero-enenrgy MBS peak is very robust against disorder and survives well 
after the SC gap in the topological phase has been suppressed to zero, and 
disappears only when the mean free path $\lambda\lesssim \xi/6$ or $E_s\gtrsim 6\Delta$ 
(for the systems are studied) whereas the topological SC gap vanishes for 
$\lambda\lesssim\xi$.

\begin{figure}[tbp]
\begin{center}
\includegraphics[width=0.47\textwidth]{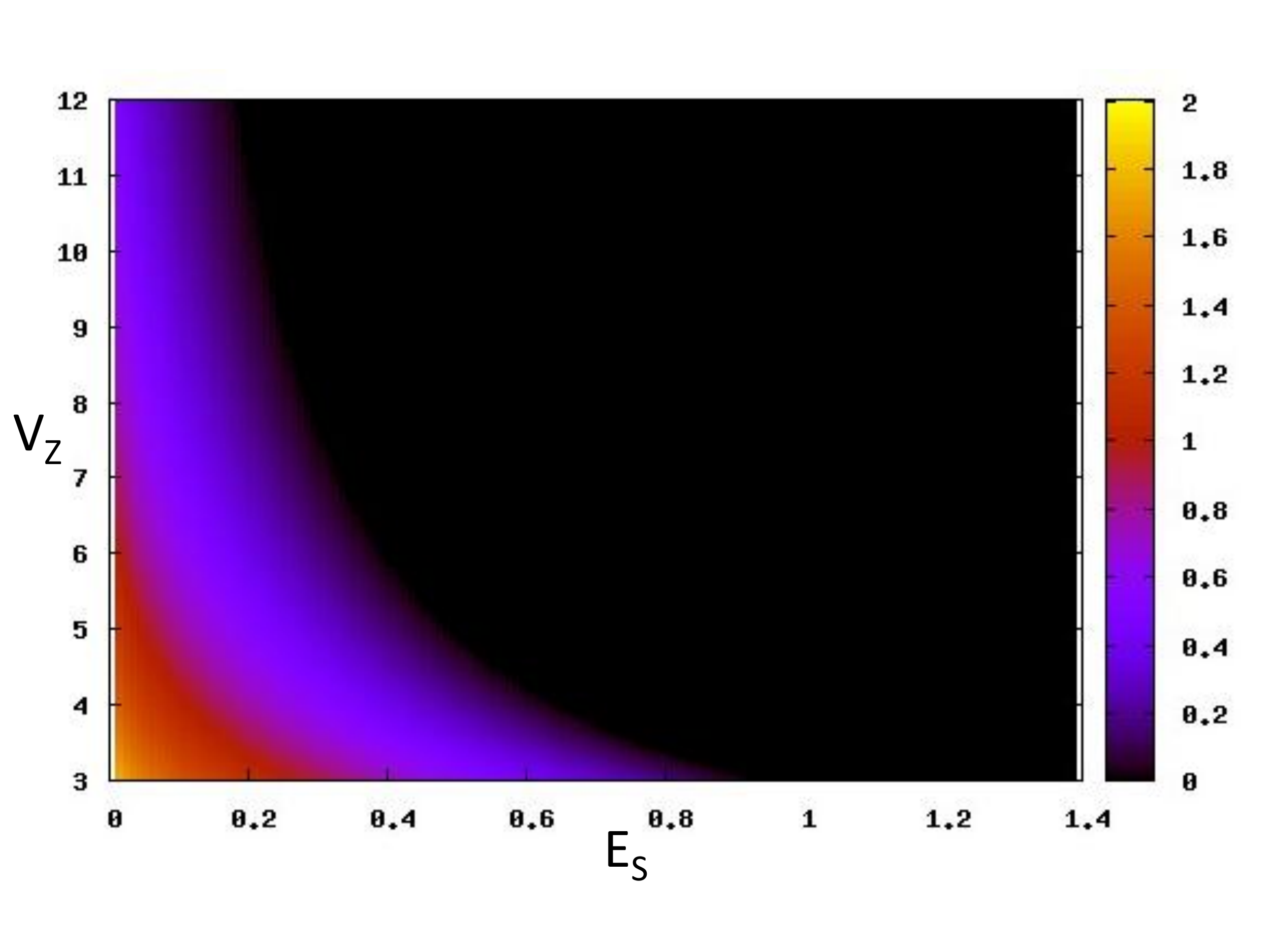}
\end{center}
\caption{Quasiparticle gap $\Delta$ (in the color bar) versus Zeeman potential $V_Z$ and scattering rate $E_s$ in the topological 
phase. The black region represents the gapless phase and is therefore not topologically robust.}
\label{Fig2}
\end{figure}

\begin{figure}[tbp]
\begin{center}
\includegraphics[width=0.7\textwidth,angle=270]{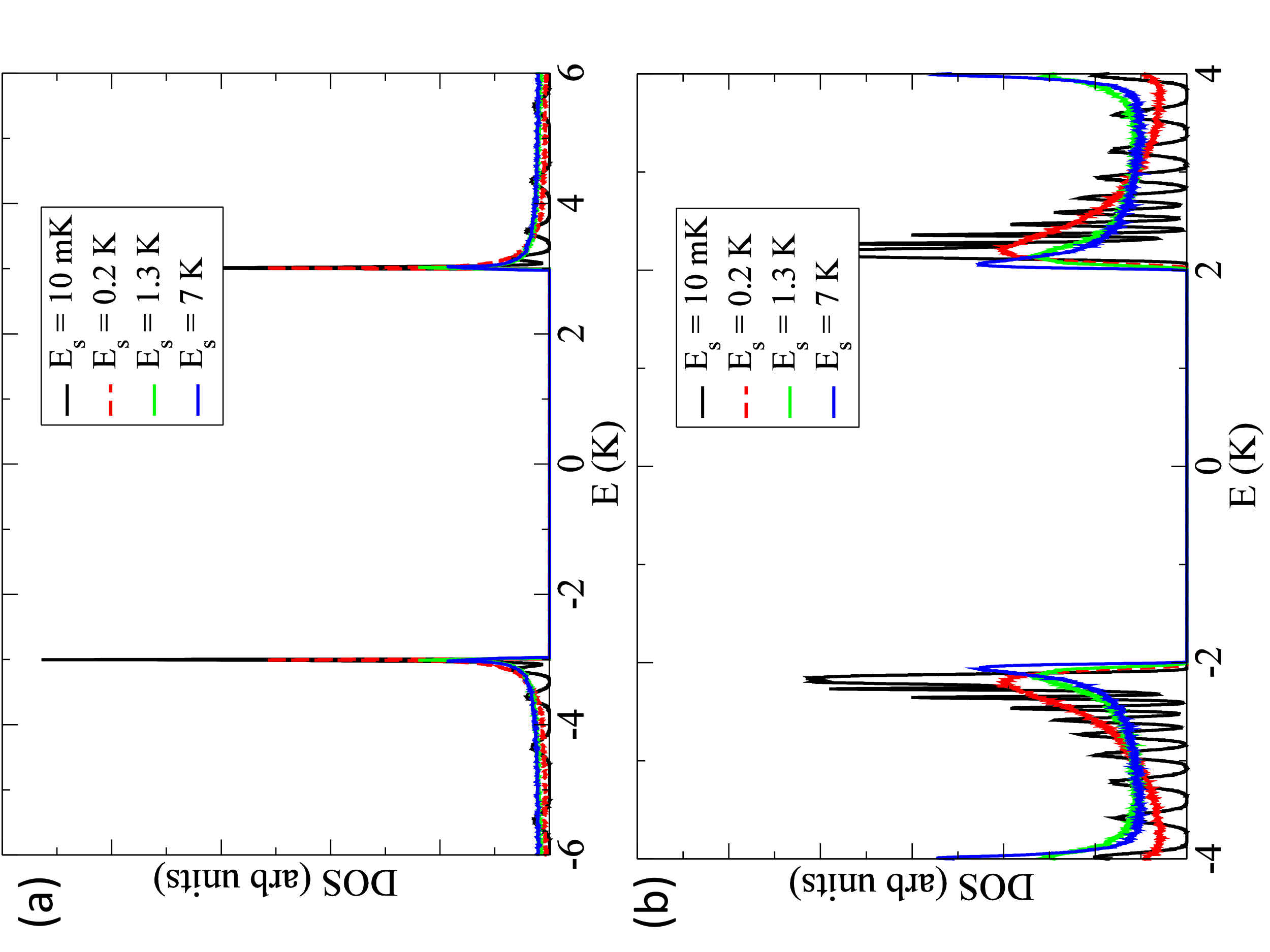}
\end{center}
\caption{Disorder averaged density of states in the non-topological phase for the semiconductor nanowire in a magnetic field with Zeeman splitting 
 $V_Z=0,\,1\,$K (panels $(a,b)$ respectively) at $\mu=0$ for different disorder strengths characterized by scattering rates $E_s=\hbar/\tau$.}
\label{Fig3}
\end{figure}

Given the quantitative validity of the SCBA, we can calculate the phase diagram 
of the SC/SM structure in the presence of disorder by using the analytical SCBA
theory. \cite{13} We show our SCBA-calculated quantum phase diagram of the system in 
Fig. 2. In Fig. 2, we fix all parameters except for disorder $(E_s)$ and 
spin-splitting $(V_Z)$ with the color representing the calculated SC gap 
$E_{SCBA}$ in the presence of disorder (by definition $E_{SCBA}=\Delta$ 
for $E_s=0$). The black region represents gapless superconducivity. 
 We note, based on our numerical results of Fig. 1, that a large part of the black
region in the phase diagram allows for well-defined zero-energy DOS MBS peaks 
although the system is essentially a gapless topological superconductor in 
this regime. Much of this SCBA gapless regime is dominated by Griffiths physics 
except for very large disorder when the system eventually becomes nontopological (i.e. 
even the finite topological segments disappear). 
Much of this SCBA gapless regime is dominated by Griffiths physics except for very large 
disorder when the system eventually becomes non-topological.
Except for our current results showing the well-defined robust 
persistence of the MBS peak even in the gapless topological regime 
(the black region in Fig. 2), one would have concluded that SCBA predicts a rather 
gloomy picture for the existence of the Majorana mode in the disordered 
SC/SM hybrid structure since, without our current exact results, the 
conclusion would have been that there cannot be any MBS in the black region 
of the phase diagram. Of course the quantitative details of the SCBA phase diagram 
depend on the SO-coupling strength and the topological phase with a large SC gap 
is easily achieved by large (small) SO coupling (disorder).

\section{DOS in the disordered non-topological phase}
Next we comment on the effect of disorder in the non-topological 
SC regime, i.e., for $V_Z<\sqrt{\Delta^2+\mu^2}$ for the SC/SM hybrid 
structure. In Fig. 3(a) and (b), we show our calculated numerical DOS in 
the non-topological SC phase (for $\mu=0$) for $V_Z=0$ (3a) and $1K$ (3b) in 
the presence of disorder. All parameters other than $V_Z$ are exactly the same 
as in Fig. 1. We see that, at least for $\mu=0$, the behavior of the DOS is 
similar to a classic $s$-wave SC (even for $V_Z=1\, K$ which simply reduces the 
gap from $\Delta_0=3\,K$ to $2\,K$) with essentially no discernible effect
on the DOS. Even for disorder as large as $E_s> 7\,K>2\Delta_0$, we do not see any structure 
developing in the SC DOS gap which remains completely flat.
For $V_Z=0$ (Fig. (3a)) this is a direct consequence of Anderson's theorem where the robustness 
of the gap arises from time-reversal symmetry. We see in Fig. (3b) that this 
behavior persists for small Zeeman fields as long as $V_Z\lesssim \Delta$. This behavior 
can be expected based on previous studies on the bound states of single impurities in 
spin-orbit coupled nanowires in proximity to superconductors \cite{27}. There it was found that the low-energy 
sub-gap states appear for short-ranged impurities only in the topological phase.  
However, this conclusion might not apply to longer-range disorder because in principle, 
at any non-zero value of Zeeman potential puddles can lead to the formation of a pair of 
low-energy MBSs. In our numerical work here we restrict ourselves only to zero-range white noise 
disorder in the semiconductor. 

\begin{figure}[tbp]
\begin{center}
\includegraphics[width=0.7\textwidth,angle=270]{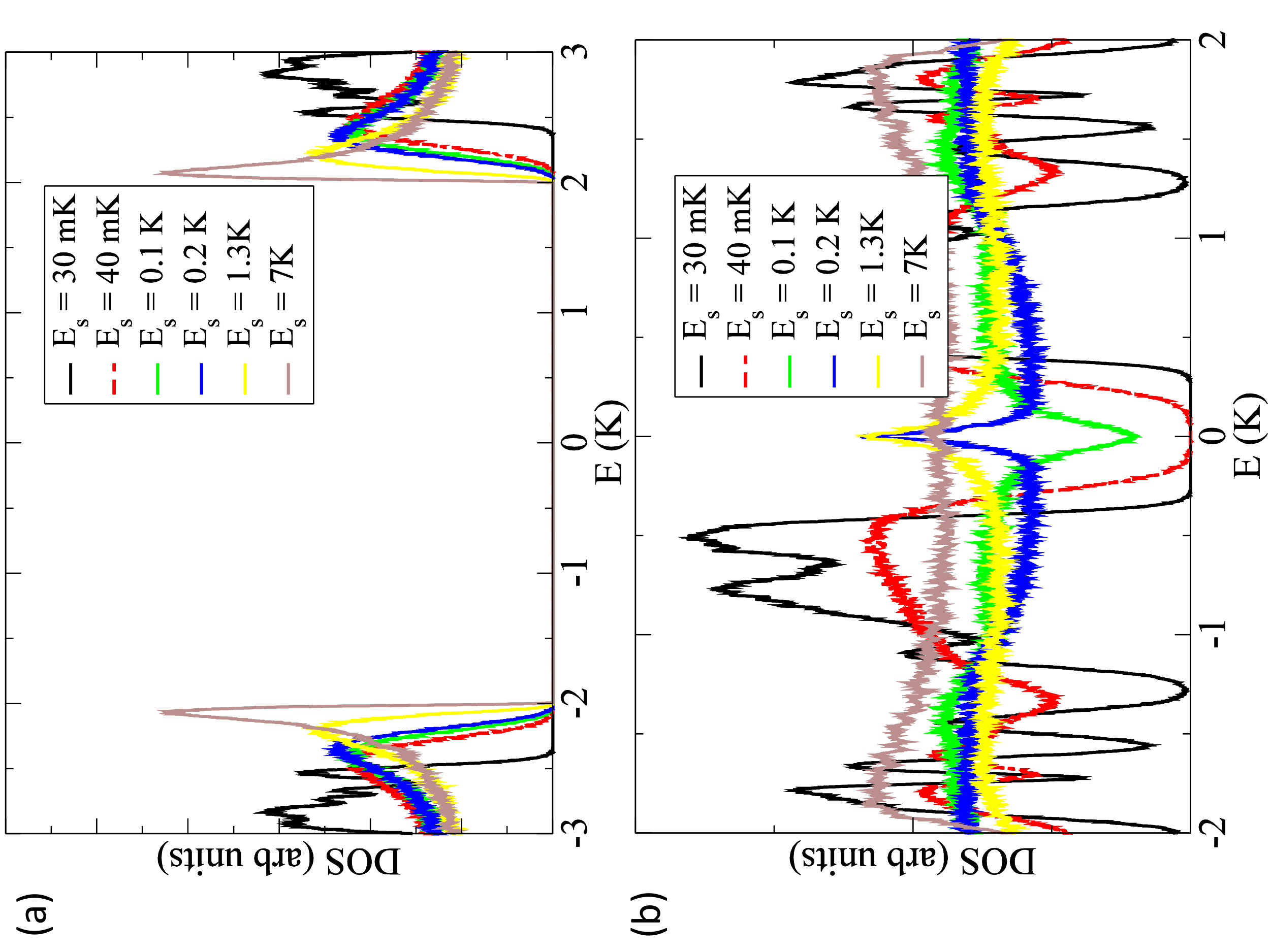}
\end{center}
\caption{Disorder averaged density of states in the non-topological phase for the semiconductor nanowire in a magnetic field with Zeeman splitting 
 $V_Z=4,\,5\,$K (panels $(a,b)$ respectively) at $\mu=5\,$K for different disorder strengths characterized by scattering rates $E_s=\hbar/\tau$.}
\label{Fig4}
\end{figure}

The situation, however, changes qualitatively when we consider the limit $V_Z\gtrsim \Delta$.
 We show these 
results in Fig. 4 where $V_Z=1\, K$, $\mu=5\, K$ and $V_Z=\mu=5\,K$ are shown for 
several values of the disorder parameter $E_s(\equiv \hbar/2\tau)$. All other 
parameters are the same as in Fig. 1 and 2. We note that the situation in Fig. 4 
describes the non-topological phase since $V_Z=\mu<\sqrt{\mu^2+\Delta^2}$.
In both panels, large disorder has a strong effect on the SC DOS shrinking the SC gap 
considerably. However, in Fig. 4(b), where $V_Z\gg\Delta$,
 eventually for $E_s=0.2\,K$ and $1.3\,K$ (and 
$V_Z=\mu=5\,K$), the DOS seems to "flip" and the dip at $E=0$ becomes 
a peak at $E=0$ with a concomitant vanishing of any SC gap feature in the 
data. For even larger $E_s$ ($E_s=7\,K$ in Fig. 4) the peak feature in the 
DOS is suppressed, but there is a clear $E=0$ DOS peak for intermediate 
(but still large with $E_s>\Delta$) disorder ($E_s=0.2\,K$ and $1.3\,K$ in 
Fig. 4) where the SC gap has disappeared, but a peak has developed in the DOS
at zero-energy. We believe that the zero-energy DOS peak in Fig. 4(b) for 
$E_s=0.2\,K$ and $1.3\,K$ has the same origin as the physics recently 
discussed in several publications \cite{16,17,18}.
  The hallmark of this "trivial" DOS peak 
(arising from the competition among spin splitting, spin-orbit coupling, 
and superconductivity) are that (1) it arises only for large Zeeman splitting  
 in the non-topological phase; (2) it 
occurs only after the SC quasiparticle gap has been completely suppressed by disorder; (3) it 
exists only in an intermediate disorder  range where the superconducting quasiparticle gap  
has been suppressed, vanishing for larger disorder and the peak becoming a 
dip (i.e. the SC gap) for smaller disorder.

\begin{figure}[tbp]
\begin{center}
\includegraphics[width=0.47\textwidth]{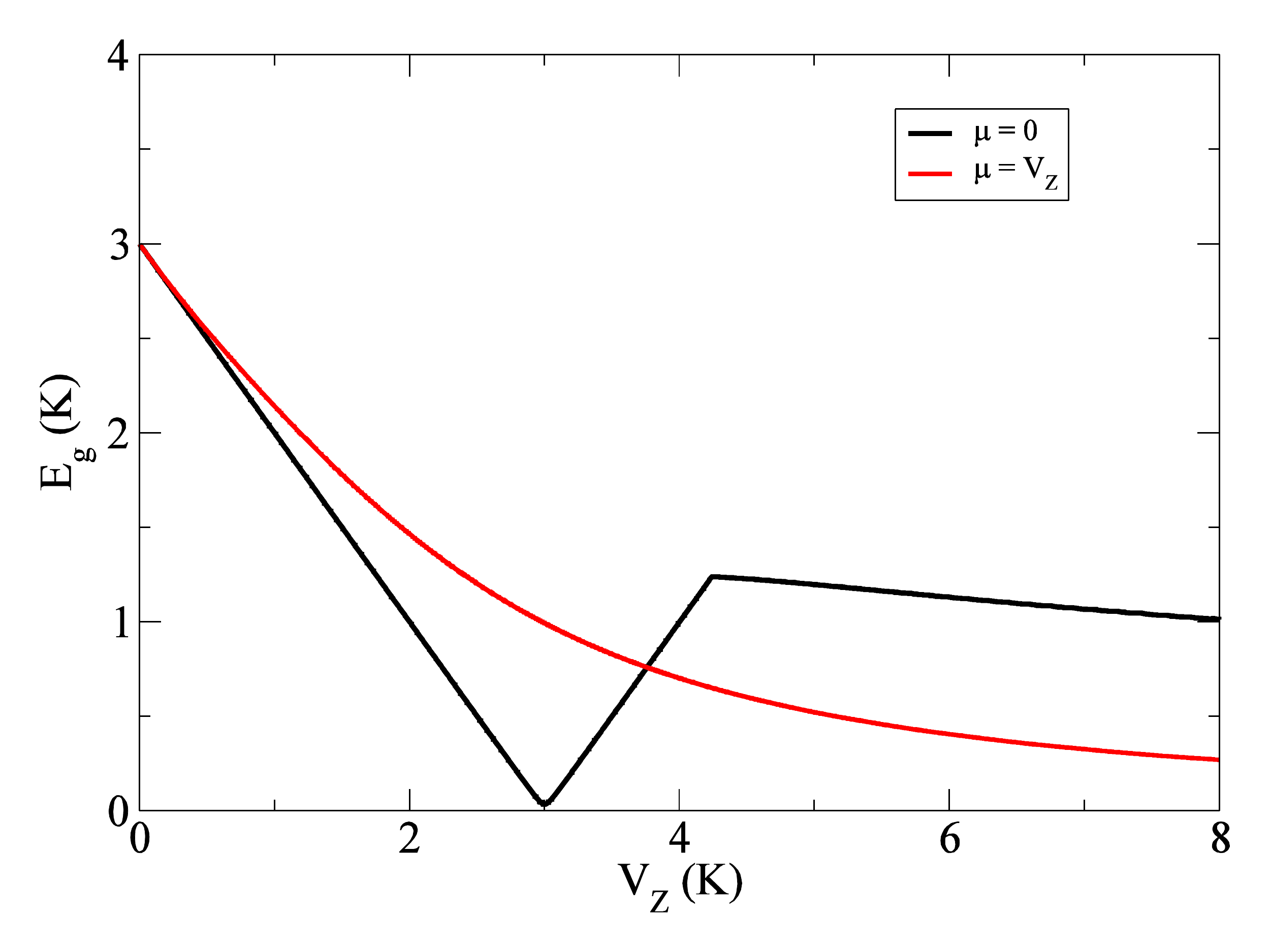}
\end{center}
\caption{Quasiparticle gap ($\Delta$) versus Zeeman potential $V_Z$ for different values ($\mu=0,\,5$K) for the chemical potential $\mu$.
The vanishing of the quasiparticle gap marks the topological quantum phase transition from the non-topological phase at small $V_Z$ to the 
topological phase at large $V_Z$.
}
\label{Fig5}
\end{figure}

In order to specify where precisely in the phase diagram we are obtaining our 
exact numerical results in the presence of disorder we finally show in Fig. 5 our 
calculated SC gap as a function of $V_Z$ (for zero disorder) for the 
parameters chosen in our calculations. For $\mu=0$ (i.e. Figs. 1-3), the 
topological quantum phase transition (TQPT) happens at $V_Z=\Delta_0(=3\,K$ 
in our choice), and our Fig. 1 belongs to the topological phase ($V_Z=5\,K$, to
the right of the TQPT) whereas our Fig. 2 belongs to the non-topological 
($V_Z=0,1\,K$) to the left of the TQPT. Results for Fig. 4 are obtained for 
$V_Z=\mu$ (and $V_Z<\mu$) which  never can manifest a TQPT since the 
$V_Z=\sqrt{\mu^2+\Delta^2}$ condition cannot be satisfied except for $\Delta=0$
 (i.e. $V_Z\rightarrow\infty$)-- but, even the nontopological SC is strongly 
affected by disorder (for $V_Z\gg \Delta$) here since the combination of spin-orbit 
coupling and Zeeman splitting makes the Anderson theorem moot.  Thus the DOS peak 
in Fig. 4, associated with anti-localization,\cite{16,17,18,19} arises purely in the trivial 
phase with a completely suppressed quasiparticle gap which might enable its experimental distinction 
from the MBS peak in the topological SC phase.

\section{Discussion}

Finally we mention that the spin-orbit coupled semiconductor system thus has two distinct 
topological quantum phase transitions in the presence of the superconducting proximity effect and 
Zeeman splitting. The first one is driven by the Zeeman field as originally predicted by Sau et al \cite{2}
with the TQPT defined by $V_Z=\sqrt{\mu^2+\Delta^2}$ assuming a low disorder situation. The second 
one is driven by increasing disorder ($E_s$) in the finite Zeeman splitting situation (i.e. $V_Z>\sqrt{\mu^2+\Delta^2}$)
where the topological SC phase is destroyed by disorder (for $E_s\gtrsim \Delta$), leading 
to a gapless non-topological phase dominated by the Griffiths physics as originally envisioned by Motrunich et al \cite{23}.
Since the effective SC gap for $V_Z\neq 0$ depends on $V_Z$, and in particular, $\Delta\propto V_Z^{-1}$ for $V_Z\gg \Delta_0$ 
[see Refs. \onlinecite{5} and \onlinecite{13} for details], we expect that there are two distinct magnetic field driven TQPTs in 
the semiconductor nanowires - - the first one is the TQPT predicted in Ref. \onlinecite{2,3,4,5} taking the system from the trivial 
$s$-wave SC (induced by proximity effect) to the (effective $p$-wave) topological Majorana carrying 
$p-$wave SC phase at low disorder (i.e. $\Delta\gg E_s$), and then the second one at much higher Zeeman splitting 
(so that $\Delta\lesssim E_s$) where disorder drives the system from a gapless topological SC to a non-topological 
Griffiths phase. It is unlikely that this second (purely disorder driven) TQPT would be experimentally accessibly since the 
gapless nature of the SC phase (i.e. black region in Fig. 2 or the situation corresponding to Fig. 1 (e)-(h)) would make the 
finite temperature of the experimental system behave like a very high temperature (i.e. $T\gg \Delta$), 
making any experimental study of this disorder-driven TQPT difficult, if not impossible. Thus, the effective gapless 
nature of the system in Fig. 1(e)-(g) would make it very unlikely that the DOS peak (which is quite obvious in our theoretical 
results) could be studied experimentally. The best hope for the direct experimental study of the MBS physics is therefore 
to have a large SC gap (as in Fig. 1(a)-(d) or in the non-black region of Fig. 2) in the topological phase which necessitates 
having low effective disorder ($E_s\ll \Delta$), a condition guaranteed by having very clean semiconductor wires and or/very strong 
SO coupling in the material. We add here that our theoretical results presented in this paper cannot be compared (or connected) with 
the experiments in any way since we only calculate bulk DOS of the system, which cannot be directly probed experimentally.

\section{Conclusion}
In summary, we have studied the disorder averaged DOS of a disordered spin-orbit coupled nanowire 
in proximity to a superconductor in both the topological and non-topological parameter regime.
The features in the DOS associated with the superconducting gap in the topological phase 
appear to be in good quantitative agreement with the SCBA from previous work \cite{13}.
Consistent with previous results \cite{13}, we find that the dips in the DOS 
associated with the quasiparticle gap disappear for relatively modest ($E_s\gtrsim \Delta$)
amounts of disorder. However, a zero-energy peak associated with MBSs generated by the Griffiths 
effect survives to much higher levels of disorder. The DOS peak associated with different levels of disorder 
arising from the Griffiths effect starting from the topological phase and the antilocalization peak starting 
from the non-topological phase appear to be different enough that one might distinguish them qualitatively. Of course, 
at this point one does not expect a sharp distinction between the peak arising from the Griffith's effect and the 
antilocalization effect because they are zero energy peaks in the DOS in the non-topological phase in the same symmetry class i.e. class D.

This work is supported by Microsoft Q, JQI-NSF-PFC, and the Harvard Quantum 
Optics Center.


\end{document}